\documentstyle[preprint,prc,aps,epsf]{revtex}
\begin{document}

\draft

\vspace{2cm}

\title{RPA quasi-elastic responses in infinite and finite nuclear systems}
\author{Eduardo Bauer \cite{AAAuth}}

\address{
Departamento de F\'{\i}sica, Facultad de Ciencias Exactas,\\
Universidad Nacional de La Plata,\\
La Plata, 1900, Argentina\\
\vskip 0.5cm
{\rm and}
}

\author{Giampaolo Co'}

\address{
Dipartimento di Fisica, Universit\`a di Lecce \\
 and Istituto Nazionale di
Fisica Nucleare, sezione di Lecce,\\
I-73100 Lecce, Italy }

\maketitle

\date{\today}

\begin{abstract}
Quasi-elastic responses in nuclear matter and in $^{12}$C and
$^{40}$Ca nuclei are calculated in ring approximation to investigate
the finite size effects on the electromagnetic quasi-elastic
responses.
A method to simulate these effects in infinite systems
calculations is proposed.
The sensitivity of the results to the various terms of the residual
interaction is studied. The results of nuclear matter RPA calculations
are compared with those obtained in ring approximation to evidence
the importance of the exchange terms.
\end{abstract}

\vspace{.5cm}

\pacs{PACS number: 21.60.Jz, 25.30.Fj, 27.20.+n, 27.40.+z}


\section{INTRODUCTION}

The quasi-elastic excitation of the nucleus is particularly
interesting for nuclear structure studies because of the interplay
between single particle and many-body effects. The excitation energies
characterizing the quasi-elastic peak are well above the nucleon
emission threshold, therefore, one or more nucleons
are ejected from the nucleus.
Before being ejected, any nucleon can interact with the others nucleons
composing the nucleus.
Considering this effect in a finite nucleus formalism is a difficult
task because the nuclear final states are described in terms of their
total angular momentum and the presence of the continuum implies the
sum on a large amount of possible configurations.

In a nuclear matter formalism one takes advantage of the translational
invariance to simplify the description of the final state of the
hadronic system. The use of the infinite system formalism is rather
appropriate for the quasi-elastic electron excitation, since excitation
energies and transferred momenta are such that the excitation process
is well localized within the nucleus and collective surface
excitations are negligible. In any case  additional approximations
(say, a variable Fermi momentum or the local density approximation)
have been considered in the literature to simulate finite size effects.

We can classify the necessary ingredients to describe the
process in three main issues:
the single particle basis,
the inclusion of initial and/or final state interactions
and the operator describing the action of the external probe.
Regarding the first point we have already said that
some work treats the nucleus as a finite
system \cite{ca84}-\cite{am94},
while other ones use a nuclear matter approach
\cite{al93}-\cite{ba00b}.

Referring to the second point, one of the
simplest approaches to include initial and final state interactions
is the Random Phase Approximation (RPA), with or without exchange
terms (the last approach named ring approximation).
In the RPA \cite{ca84}-\cite{co88},
\cite{al93}-\cite{ba99}
one particle-one hole ($1p1h$) excitations
are summed up to infinite order. One step further in complexity is
the so-called Second RPA (SRPA) \cite{dr90}, \cite{ba00} which, in
addition to the $1p1h$ excitations,
considers also those generated by $2p2h$. The
Extended RPA (ERPA) \cite{ta88}, \cite{al84}, \cite{ba95};
contains ground state correlations beyond RPA.
The Green function approach of Ref. \cite{ch89} is based
on a philosophy similar to that of the SRPA. In this approach
the relationship between forward virtual Compton scattering and
inclusive electron scattering is
used to construct a one-body approximation to quasi-elastic
electron scattering.
The role of the short-range correlations has been investigated in the
framework of the correlated basis function theory
\cite{co00},\cite{fa84}.

Concerning the third point, the external operator is usually
represented by electromagnetic one-body operators, but
in the transverse channel it is important to include two-body terms
as the Meson Exchange Currents (MEC) \cite{am93}, \cite{am94},
\cite{or81}, \cite{ba00b}, and the excitation of the virtual or real
 $\Delta(1232)$ resonance \cite{gi97}-\cite{am99}.

In the present paper we
investigate the sources of some inconsistencies
between finite and infinite nucleus calculations.
For example it seems that, once the residual interactions has been fixed,
the effects of the ring approximation are larger in nuclear matter
than in finite nuclei.
To simulate the finite size effects in the quasi-elastic peak,
we have done infinite systems calculations with a diffused
Fermi surface. Then, we have studied the effects on the responses
of the various characteristics of the interaction,
such as the different channels, the range and the density
dependence.
The infinite systems results have been compared with the calculations
done for the $^{12}$C and $^{40}$Ca nuclei, two doubly magic
nuclei with the same number of protons and neutrons.
The agreement we have finally obtained between the two kinds of
calculations is satisfactory,
even for the nucleus $^{12}$C, which is supposed to
be relatively light to be well represented as an infinite system of
nucleons.

Encouraged by this result, we have evaluated the response functions
within the nuclear matter RPA framework of Ref. \cite{ba96} to
investigate the effects of the exchange terms neglected in the ring
approximation.

The paper is organized as follows. In Sec. II we briefly present
the effective theories used to perform the calculations. In Sec. III
we compare the results obtained in finite nuclei and
nuclear matter both for the free and the ring responses, and
we present the nuclear matter RPA responses.
Finally, in Sec. IV we draw our conclusions.

\section{FORMALISM}
\label{FORM}
The response function for inclusive quasi-elastic electron scattering
is given by,
\begin{equation}
\label{RESP}
\displaystyle
R({\mbox{\boldmath $q$}}, \omega) = -\frac{1}{\pi}\ Im
<0|{\cal O}^{\dag} ({\mbox{\boldmath $q$}})
G( \omega) {\cal O}({\mbox{\boldmath $q$}}) |0>
\end{equation}
where $\omega$ represents the excitation
energy and ${\mbox{\boldmath $q$}}$  the
three-momentum transferred by the electron.
The nuclear ground state is denoted as $|0>$, while
${\cal O}({\mbox{\boldmath $q$}})$ is the excitation operator and
$G(\omega)$ the polarization propagator,
\begin{equation}
\label{GREEN}
\displaystyle
G(\omega) = \frac{1}{ \omega - H + i \eta}\ -
\frac{1}{ \omega + H + i \eta}\
\end{equation}
where $H$ is the nuclear Hamiltonian.
We introduce the
operator $P$, which projects onto $1p1h$ configurations.
The way of building these configurations depends on the particular
choice of the single particle basis.
In nuclear matter the single particle wave functions are plane waves,
while in a finite nucleus they are eigenfunctions of the one-body
Schr\"odinger equation for a mean field potential. In our case we use
a real, spherical, Woods-Saxon potential.

The response function generated by $1p1h$ excitations can be expressed
as:
\begin{equation}
\label{RESPP}
\displaystyle
R_{PP}({\mbox{\boldmath $q$}}, \omega) = -\frac{1}{\pi}\ Im
<0|{\cal O}^{\dag}({\mbox{\boldmath $q$}})
P G(\omega) P
{\cal O} ({\mbox{\boldmath $q$}})|0>
\end{equation}

The evaluation of $R_{PP}$ is not straightforward  since the
nuclear Hamiltonian is in general not diagonal in the $1p1h$ basis.
The solution of the problem is the RPA response. In the present
work, we shall compare finite nucleus results obtained within the ring
approximation of Ref. \cite{co88}, with the corresponding
nuclear matter responses.

In most calculations done in infinite systems the Fermi surface which
separates, in momentum or energy space, the hole from the particle
states, is a sharp step function.
In the next section, we shall show that this approximation
reproduces rather well the position and the
shape of the free responses calculated in finite nuclei.
However, the ring responses are appreciably different
when evaluated in nuclear matter and in finite nuclei.
The original motivation of the present work was to explain this
discrepancy. The first attempt to improve our nuclear matter model,
is to replace the step function representing the momentum
distribution of particles and holes, by a more realistic one.
For the holes with momentum ${\mbox{\boldmath $h$}}$
we make the substitution,
\begin{equation}
\label{STEP}
\displaystyle
\theta(k_F-|{\mbox{\boldmath $h$}}|) \; \rightarrow \; n(h)
\end{equation}
where $k_F$ is the Fermi momentum,
$h \equiv |{\mbox{\boldmath $h$}}|$, and
\begin{equation}
\label{DIFF} \displaystyle n(h)= \frac{1}{1+e^{(h-k_F)/a}}
\end{equation}
being $a$ a constant to be adjusted.
An analogous expression can be obtained for particles.

In the next section we shall compare
results obtained with this smoothed Fermi surface in ring
approximation with those obtained in finite systems calculations.
With the same smoothed momentum distribution we have done
RPA calculations following the computational scheme developed
in Ref. \cite{ba96} which we briefly describe.
In nuclear matter, direct  RPA  terms can be summed up to infinite
order (ring series), but, in general, it is not
possible to find a closed form to sum all the exchange terms.
Normally these last terms are perturbatively considered and,
for numerical reasons, this is done up to the second order.
On the other hand, it possible to make the full summation
of the exchange terms when a contact interaction is used. Full sums
are also possible for separable interactions.
To exploit this feature we rewrite the residual interaction $V$, as:
\begin{equation}
\label{VRES}
\displaystyle
V = V_{contact} + \tilde{V}
\end{equation}
The $V_{contact}$ term is a contact interaction conveniently
chosen to make $\tilde{V}$ small.
For $V_{contact}$ both direct and exchange terms are
summed up to infinite order while $\tilde{V}$ is perturbatively
considered up to the second order which we found to account
reasonably well for the whole sum.
Interference terms between $V_{contact}$ and $\tilde{V}$ are
included up to infinite order in $V_{contact}$ and up to second
order in $\tilde{V}$. A more detailed description of the method can
be found in Ref. \cite{ba96}.

The finite nucleus calculations have been done within the
Fourier-Bessel computational scheme adopted in Ref. \cite{co88}. The
single particle basis is generated by using a Woods-Saxon mean field
whose parameters have been fixed to reproduce experimental rms radii
and single particle energies of the bound states close to the Fermi
surface. By neglecting the exchange diagrams, the RPA equations are
rewritten in terms of local density functions which are expanded on a
Fourier-Bessel basis. In this manner the problem to be solved, for
every value of the excitation energy, is the diagonalization of a
matrix whose dimensions are four times the number of the
Fourier-Bessel expansion coefficients. More details about the method
can be found in Ref. \cite{ha82}.

\section{RESULTS}
In this section we compare finite nucleus  results with
those of nuclear matter. This is done for two nuclei:
$^{12}$C and $^{40}$Ca. The $^{12}$C nucleus is perhaps too light
to be appropriately described in terms of nuclear matter.
On the other hand we wanted to test our model also in extreme
situations and, last but not least, the finite nucleus calculations
are much less involved than in the $^{40}$Ca case.

The transferred momenta
analyzed are $q=400$ MeV/c and $q=500$ MeV/c.
These values are sufficiently large to eliminate the presence of
collective surface vibrations, and at the same time, sufficiently small
to require a limited number of partial waves. In the finite nucleus
calculations we sum multipole excitations up to angular momentum J=12
\cite{am93}.

The first step  of our calculations consists in fixing the values of
$k_F$ and $a$ in eq. (\ref{DIFF}) to
reproduce the finite nucleus free responses.
For $^{12}$C we have obtained the values $k_F$=0.85 fm$^{-1}$  and
$a$=0.20 fm$^{-1}$ ,
and for $^{40}$Ca, $k_F$=1.0 fm$^{-1}$ and $a$=0.17
fm$^{-1}$.
The comparison with the finite nuclei responses is shown
in Figs. \ref{fig:free12}
and \ref{fig:free40}, where we have also added the results obtained
with a step function Fermi surface represented by the dashed
lines.
In this last case we have used the procedure of Ref. \cite{am94} to
fix the value of the Fermi momentum
obtaining the values $k_F$=1.09  fm$^{-1}$  for $^{12}$C and
$k_F$=1.19  fm$^{-1}$  for $^{40}$Ca. These values are noticeably
different from those fixed by the smoothed momentum distribution.
As expected, $k_F$ increases with increasing mass number.
We succeeded in obtaining a satisfactory
agreement with the finite nucleus responses,
especially for $^{40}$Ca, in the case of a diffused Fermi surface.
As expected the high energy tail of the
finite nucleus responses can be reproduced only by the calculations
with  a diffused Fermi surface.

The comparison between the various responses calculated in ring
approximation is shown in Figs. \ref{fig:ringpp12} and
\ref{fig:ringpp40} for the $^{12}$C and $^{40}$Ca nuclei respectively.
In these figures the meaning of the symbols is analogous to that of
the previous figures. The interaction used in these calculations is the
finite range polarization potential utilized also in Ref. \cite{co88}.
The nuclear matter calculations have been done with the values of
$k_F$ and $a$ previously fixed.

Two observations should be done about these results. A first one is
about the fact that, in general, the full curves reproduce better the
finite nucleus results than the dashed ones. In a second place
we observe that the longitudinal responses are better reproduced
than the transverse ones.

The first observation induces to conclude that the sharp Fermi surface,
even with an effective value of the Fermi momentum, is unable to
reproduce the finite nucleus results. This feature does not depend
from the residual interaction, as we show in
Fig. \ref{fig:ringmi12} where the various  $^{12}$C responses have
been calculated, always in ring approximation, with a zero range
Migdal interaction.
We have used the following values of the parameters of this force:
$f_0=386$, $f'_0=289.5$, $g_0=106.2$ and $g'_0=135.1$,
expressed in MeV fm$^3$ units. These values have been chosen to
magnify some of the effects we want to discuss. Specifically,
the big value of $f_0$, ten times larger than the one normally used
\cite{ri78}, enhances the difference
between the nuclear matter calculations in the longitudinal
response. Here it becomes more evident the poor quality of the sharp
Fermi surface calculations. Our diffused Fermi surface calculation is
able to reproduce rather well the finite nucleus results even in these
extreme conditions.

Concerning the observation that, the longitudinal responses are
always better reproduced than the transverse ones, we have verified
that this fact is due to the differences between infinite and finite
systems. The isospin channel of the force does not contribute in ring
approximation calculations of nuclear matter transverse responses,
while it does in finite nuclei calculations. The effects of this
difference become evident by comparing the transverse responses of
Fig. \ref{fig:ringmi12} with Fig. \ref{fig:ringf112}.
In this last figure
the $^{12}$C transverse responses have been calculated with the
contact interaction above described but without the isospin channel of
the force, i.e. by setting $f'_0=0$. The agreement between
the finite nucleus responses and those obtained with the diffused
Fermi surface is comparable with that obtained in the longitudinal
case.

We should remark the fact that the polarization
potential has a density dependence in the scalar and isospin
channels. In the nuclear matter calculations we have used the force
parameters defined for the nuclear interior. We have checked the
sensitivity of the results by switching off the density dependence in
the finite nuclei calculations and comparing with the responses
evaluated with the full interactions. The differences found between
these two calculations are of the same order of the differences
between nuclear matter and finite nuclei longitudinal responses. We
conclude that for the calculations of quasi-elastic responses the
density dependence of the force is not important.

In Figs. \ref{fig:rpapp12} and \ref{fig:rpapp40} we compare
the RPA nuclear matter responses (full lines) with those evaluated in ring
approximation (dashed lines).
Both calculations have been done by using the polarization potential
and the diffused Fermi surface. This comparison shows the effects
of the exchange diagrams evaluated in RPA and neglected in ring
approximation. For the particular interaction used
these effects are noticeable,
especially in the longitudinal responses and for low values of the
momentum transfer.

In the same figures we also present the free responses (dotted lines)
and the $^{12}$C and $^{40}$Ca experimental points.
Our results show that the major source of disagreement between free
responses and data is produced by correlations beyond the RPA.
In Ref. \cite{co88} the role of the final state
interactions and of the effective mass was pointed out.
The inclusion of these two effects
within a simplified model produces a good agreement with the
$^{40}$Ca data \cite{am93}. The same model is however unable to
explain the $^{12}$C transverse response data.

\section{CONCLUSIONS}
In the present work we have compared nuclear matter and finite nuclei
quasi-elastic responses induced by electron scattering.
We have proposed a computational scheme which is able to reproduce the main
features of the finite nucleus and, at the same time, it has the
computational advantages of nuclear matter.
This has been achieved by using a diffused momentum distribution
of particles and holes in nuclear matter calculations.
We have described the momentum distribution with a simple Fermi
function depending from two parameters whose values have been adjusted
to reproduce the finite nuclei free responses.

With this simple model we have calculated the $^{12}$C and $^{40}$Ca
responses for different values of the momentum transfer in ring
approximation. The agreement with the finite nucleus calculations
is excellent for the longitudinal
responses. We traced the source of the
small differences found in the transverse responses to the isospin
part of the residual interaction which is not active in nuclear
matter calculations.

Using the same residual interaction, the polarization potential, we have
done RPA calculations, i. e. we have also considered the exchange
diagrams. The aim of this calculation was to test its
feasibility. Finite nuclei RPA calculations have been done
\cite{ca84}-\cite{bu91},
but they require a large computational effort, while
our method is simpler and more suitable to be used for
those extensions beyond RPA which
are necessary to describe the experimental points.

With respect to this last point some words of caution are
necessary to avoid double counting. A first one is about the residual
interaction which in RPA calculations cannot be the polarization
potential constructed in Ref. \cite{pi88} to consider in
average manner the exchange diagrams. A second warning is about the
diffuseness of the Fermi surface which is partially produced
by correlations effects beyond RPA.

Our approach should be compared with the most commonly used method to
account for finite nucleus effects in nuclear matter
calculations: the local density approximation (LDA).
Unfortunately this comparison is not straightforward.
In LDA the responses are calculated for several values of the
Fermi momentum and then they are appropriately averaged to obtain
the final result.
In the averaging procedure
the way how the responses for each $k_F$ are weighted, differs
from nucleus to nucleus. However, for each $k_F$ a sharp Fermi
surface is employed and this implies that the second term in the right
hand side of Eq. (\ref{GREEN}) does not contribute.
This terms, however, gives a contribution when a smooth Fermi
surface is considered. This contribution is normally small but it
becomes appreciable when $\omega \leq $ 20 MeV .
This is the  formal difference between our method and the LDA.
From the pragmatical point of view the poor quality of LDA results in
reproducing finite nuclei quasi-elastic responses has already
been pointed out
in Ref. \cite{am94}, where the effective momentum approximation was
found to be better.
Here we have shown the superiority of the diffused Fermi
surface scheme even with respect to the effective Fermi momentum
approximation. One should also consider that also from the
computational
point of view our scheme is superior to the LDA. In the last case one
has to calculate nuclear matter responses for different values of
$k_F$, while in our case a single calculation is necessary.
For the fast calculations done in ring approximation, this
difficulty is irrelevant, but it becomes an handicap for
RPA calculations, or for more elaborated ones like SRPA or ERPA.

\acknowledgements
We thank A.M. Lallena for useful discussions.


%
%
%
%
%
\begin{figure}
\caption{ Nuclear matter free responses
  for two values of
  the momentum transfer compared with the continuum shell model
  $^{12}$C responses (black points). The full lines have been obtained
  using a diffused Fermi surface, the dashed ones with a step
  functions Fermi surface. In both cases the values of the parameters
  have been modified to reproduce the finite nucleus responses.
}
\label{fig:free12}
\end{figure}
%
%
%
\begin{figure}
\caption{ The same as Fig. \ref{fig:free12} but for  $^{40}$Ca. }
\label{fig:free40}
\end{figure}
%
%
%
\begin{figure}
\caption{  $^{12}$C responses calculated in ring approximation with
  the polarization potential. The meaning of the lines and of the
  symbols is the same as in Fig. \ref{fig:free12}.
}
\label{fig:ringpp12}
\end{figure}
%
%
%
\begin{figure}
\caption{  Same as Fig. \ref{fig:ringpp12} for $^{40}$Ca. }
\label{fig:ringpp40}
\end{figure}
%
%
%
\begin{figure}
\caption{  $^{12}$C responses calculated in ring approximation
  with the zero-range Migdal interaction. }
\label{fig:ringmi12}
\end{figure}
%
%
\begin{figure}
\caption{  Transverse $^{12}$C responses calculated in ring
  approximation with a contact interaction. The
  spin-isospin term was left out.
}
\label{fig:ringf112}
\end{figure}
%
%
\begin{figure}
\caption{  $^{12}$C responses calculated in RPA (solid lines) and
  ring approximation (dashed lines) with
  the polarization potential. The dotted lines show the free responses.
  Data from ref. \protect\cite{ba83}
}
\label{fig:rpapp12}
\end{figure}
%
%
\begin{figure}
\caption{  The same as Fig. \ref{fig:rpapp12} but for $^{40}$Ca.
 Data from ref. \protect\cite{wi97}.
}
\label{fig:rpapp40}
\end{figure}

\begin{references}

\bibitem[*]{AAAuth}Fellow of the Consejo Nacional
de Investigaciones Cient\'{\i}ficas y T\'ecnicas, CONICET.

\bibitem {ca84}
M. Cavinato, D. Drechsel, E. Fein, M. Marangoni and A.M. Saruis,
Nucl. Phys. {\bf A423}, 376 (1984);
M. Cavinato, M. Marangoni and A.M. Saruis, Phys. Lett. {\bf B235}, 346
(1990); A.M. Saruis, Phys. Rep. {\bf 235}, 57 (1993).
%
\bibitem{de85}
A. Dellafiore, F. Lenz and F.A. Brieva,
Phys. Rev. {\bf C31}, 1088 (1985);
F.A. Brieva and A. Dellafiore, Phys. Rev. {\bf C36}, 899 (1987).

\bibitem {sh89}
T. Shigehara, K. Shimizu and A. Arima, Nucl. Phys. {\bf A492}, 388
(1989).

\bibitem {bu91}
M. Buballa, S. Dro\.zd\.z, S. Krewald and J. Speth, Ann. Phys. (N.Y.)
{\bf 208}, 346 (1991);
M. Buballa, S. Dro\.zd\.z, S. Krewald and A. Szczurek, Phys. Rev. C
{\bf 44}, 810 (1991);
S. Jeschonnek, A. Szczurek, G. Co' and S. Krewald, Nucl. Phys.
{\bf A570}, 599 (1994).

\bibitem  {co88}
G. Co', K. F. Quader, D. R. Smith and J. Wambach,
Nucl. Phys. {\bf A485}, 61 (1988).

\bibitem  {dr90}
S. Dro\.zd\.z, G. Co', J. Wambach and J. Speth,
Phys. Lett. {\bf B185}, 287 (1987);
S. Dro\.zd\.z, S. Nishizaki, J. Speth and J. Wambach,
Phys. Rep. {\bf 197} 1 (1990).

\bibitem  {ta88}
K. Takayanagi, K. Shimizu and A. Arima,
Nucl. Phys. {\bf A477}, 205 (1988);
K. Takayanagi, Phys. Lett. {\bf B230}, 11 (1989);
 Nucl. Phys. {\bf A510}, 162 (1990);
 {\it ibid.} {\bf A522}, 494 (1991);
 {\it ibid.} {\bf A522}, 523 (1991);
 {\it ibid.} {\bf A556}, 14 (1993).

\bibitem  {ch89}
C. R. Chinn, A. Picklesimer and J. W. Van Orden,
Phys. Rev. {\bf C40} 1159 (1989);
F. Capuzzi, C. Giusti and F. D. Pacati,
Nucl. Phys. {\bf A524} 681 (1991);
F. Capuzzi, Nucl. Phys. {\bf A554} 362 (1993).

\bibitem {am93}
J. E. Amaro, G. Co' and A. M. Lallena, Ann. Phys. (N.Y.) {\bf 221},
306 (1993); Nucl. Phys. {\bf A578}, 365 (1994).

\bibitem{co00}  G. Co' and A. M. Lallena, Ann. Phys. (N.Y.) (to be
  published), arXiv:nucl-th/0007032.

\bibitem {am94}
J. E. Amaro, A.M. Lallena and G. Co', Int. J. Mod. Phys.
{\bf E3}, 735 (1994).

\bibitem {al93}
W. M. Alberico, M. B. Barbaro, A. De Pace, T. W. Donnelly and A. Molinari,
Nucl. Phys. {\bf A563}, 605 (1993);
M. B. Barbaro, A. De Pace, T. W. Donnelly and A. Molinari,
Nucl. Phys. {\bf A596}, 553 (1996); {\it ibid.} {\bf A598}, 503 (1996);
A. De Pace, Nucl. Phys. {\bf A635}, 163 (1998).

\bibitem {ba96}
E. Bauer, A. Ramos and A. Polls, Phys. Rev. {\bf C54}, 2959 (1996).

\bibitem {ba99}
E. Bauer and A. Lallena, Phys. Rev. {\bf C59}, 2603 (1999).

\bibitem {ba00}
E. Bauer, J. Phys. {\bf G26}, 145 (2000).

\bibitem{al84}
W. M. Alberico, M. Ericson and A. Molinari, Ann. Phys. (N.Y.)
{\bf 154}, 356 (1984);
W. M. Alberico, A. De Pace, A. Drago and A. Molinari, Rivista del
Nuovo Cimento {\bf 14}, 1 (1991);
W. M. Alberico, T. W. Donnelly and A. Molinari, Nucl. Phys. {\bf A512},
541 (1990).

\bibitem {ba95}
E. Bauer, Nucl. Phys. {\bf A589}, 669 (1995);
E. Bauer, A. Polls and A. Ramos, Phys. Rev. {\bf C58}, 1052 (1998).

\bibitem  {fa84}
S. Fantoni and V. R. Pandharipande, Nucl. Phys. {\bf A427}, 473 (1984);
A. Fabrocini and S. Fantoni, Nucl. Phys. {\bf A503}, 375 (1989);
O. Benhar,  A. Fabrocini and S. Fantoni,
Nucl. Phys. {\bf A550}, 201 (1992).

\bibitem {gi97}
A. Gil, J. Nieves and E. Oset, Nucl. Phys. {\bf A627}, 543 (1997);
A. Gil, Ph. D. Thesis, Universitat de Val\`encia, 1996.

\bibitem{ce97}
R. Cenni, F. Conte and P. Saracco, Nucl. Phys. {\bf A623}, 391 (1997).

\bibitem {ba98}
E. Bauer, Nucl. Phys. {\bf A637}, 243 (1998).

\bibitem {am99}
J. E. Amaro, M. B. Barbaro, J. A. Caballero, T. W. Donnelly
and A. Molinari,
Nucl. Phys. {\bf A657}, 161 (1999).

\bibitem  {or81}
J. W. Van Orden and T. W. Donnelly,
Ann. Phys. (N.Y.) {\bf 131}, 451 (1981).

\bibitem{ha82}
R. de Haro, S. Krewald and J. Speth,
Nucl. Phys. {\bf A388}, 265 (1982);
G. Co' and S. Krewald,
Nucl. Phys. {\bf A433}, 392 (1985).

\bibitem {ba00b}
E. Bauer, Phys. Rev. {\bf C61}, (2000).

\bibitem  {ri78}
G. Rinker and J. Speth,
Nucl. Phys. {\bf A306}, 360 (1978).

\bibitem  {ba83}
P. Barreau {\it et al.},
Nucl. Phys. {\bf A402}, 515 (1983).

\bibitem {wi97}
C. F. Williamson {\it et al.},
Phys. Rev. {\bf C56}, 3152 (1997).

\bibitem {pi88}
D. Pines, K.F. Quader and J. Wambach,
Nucl. Phys. {\bf A477}, 365 (1988).

\end{references}
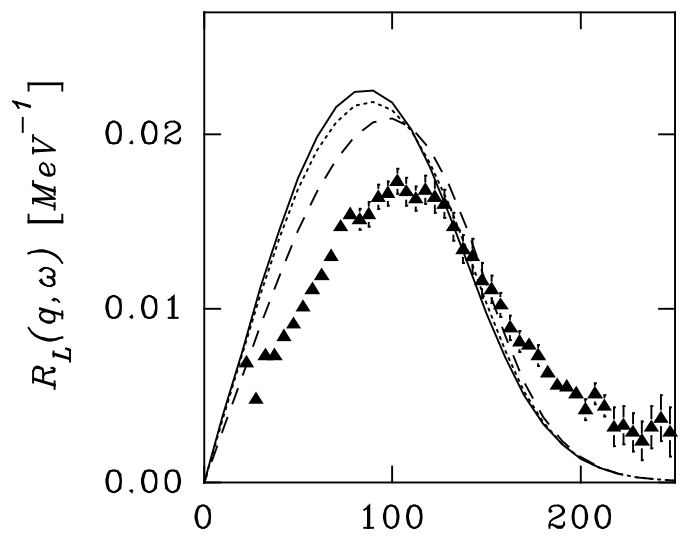
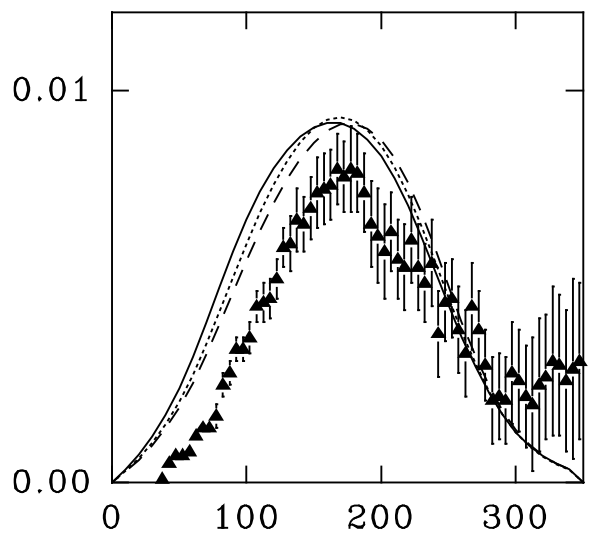
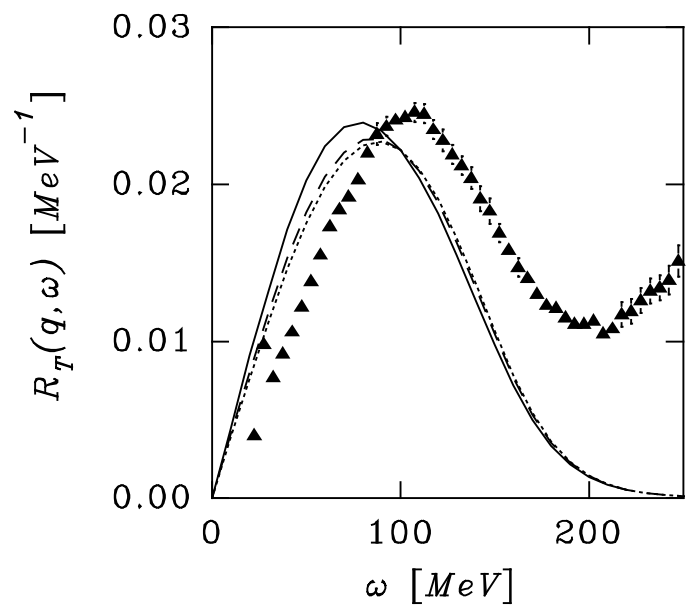
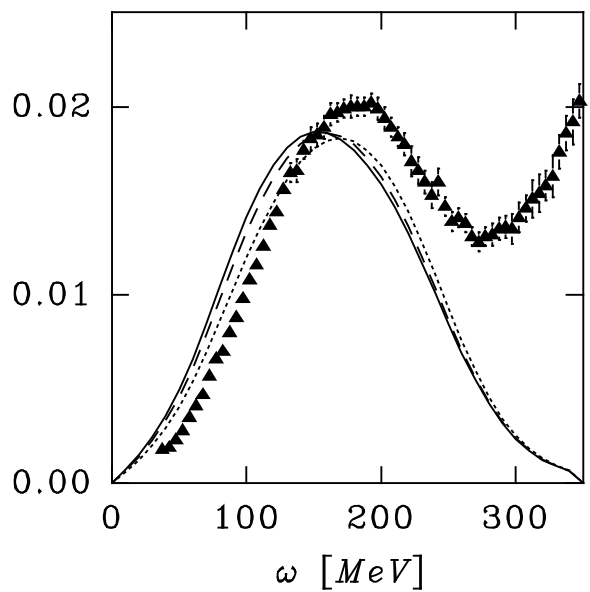
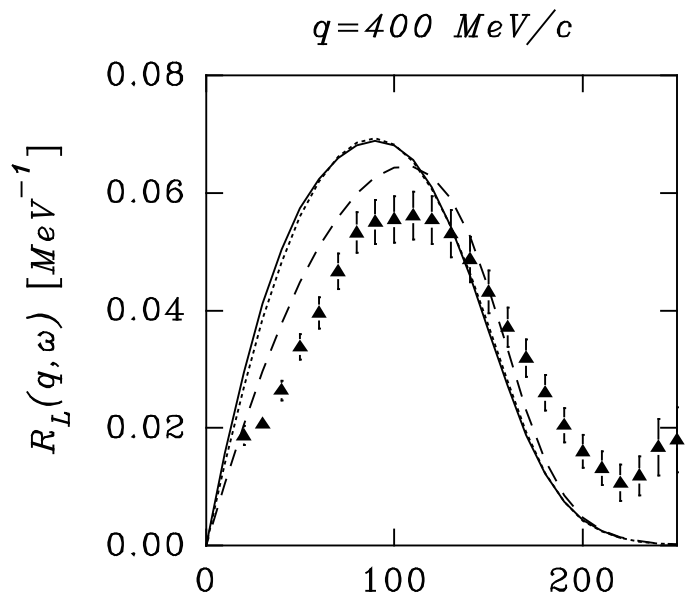
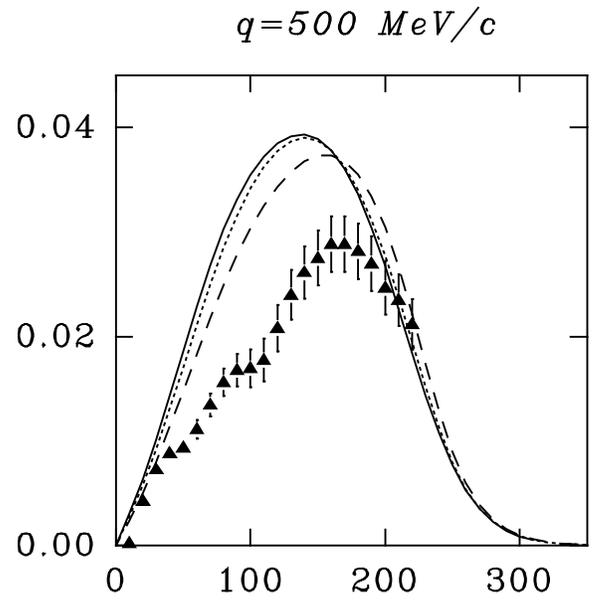
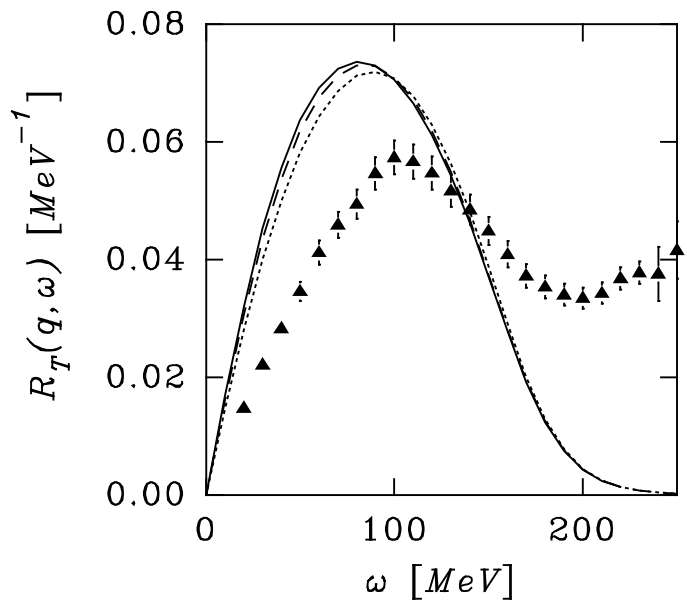
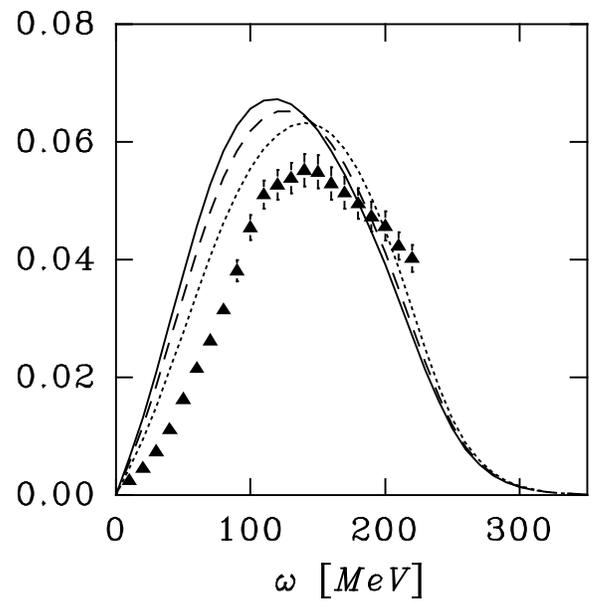
\end{document}